\documentclass[conference]{IEEEtran}
\usepackage{graphicx}

\usepackage{multirow}
\usepackage{url}

\begin{document}

\title{Activity Detection from Encrypted Remote Desktop Protocol Traffic}

\author{\IEEEauthorblockN{L. Lapczyk}
\IEEEauthorblockA{Centre for Advanced Computing\\
Queen's University\\
Kingston, Canada\\
Email: 17ll16@queensu.ca}
\and
\IEEEauthorblockN{D.B. Skillicorn}
\IEEEauthorblockA{School of Computing\\
	Queen's University\\
	Kingston, Canada\\
	Email: skill@cs.queensu.ca}
}

\maketitle

\begin{abstract}
An increasing amount of Internet traffic has its content encrypted.
We address the question of whether it is possible to predict
the activities taking place over an encrypted channel, in
particular Microsoft's Remote Desktop Protocol.
We show that the presence of five typical activities can be
detected with precision greater than 97\% and recall greater than 94\%
in 30-second traces.
We also show that the design of the protocol exposes fine-grained
actions such as keystrokes and mouse movements which may be
leveraged to reveal properties such as lengths of passwords.
\end{abstract}

\section{Introduction}

One response to the obvious security weaknesses of networks is to
encrypt the payloads of packet traffic.
It was estimated in 2018 that encryption was used
in more than 70\% of network communications \cite{maddison2018} 
but the rate of penetration is quite variable because of the
cost and complexity of public key infrastructure, particularly for
small enterprises.
We explore the question of how much can be detected about
interactions, even when payloads are encrypted.
In particular, we are concerned with activity detection: what
is a user doing?

There are some legitimate reasons to be able to answer this question:
determining usage to insure adequate provisioning, for example.
It is also important to know how much an adversary could infer
\cite{zhang2018homonit}. 
We focus on the Microsoft Remote Desktop Protocol (RDP), a
popular protocol that provides an encrypted channel between a client
and a host, allowing remote work on the host \cite{dautis_2018}. 
Five major interactions occur between client and host: file
download, browsing on the host (i.e. a mixture of viewing, typing,
and using the mouse), using an editor on the host, watching a video,
and copying content from host to client or \emph{vice versa}
using the clipboard.

We show that, even though all traffic is encrypted, it is possible to
detect which of these activities is underway from the traffic properties,
even if two or more simultaneous activities are occurring.
A heterogeneous ensemble classifier achieves
precision greater than 97\% and recall greater than 94\%.
We also discover that there are markers
in the protocol that are directly related to fine-grained user actions. 

\section{Related work}

Traffic inspection and classification has two main purposes: 
security (detecting malicious activity \cite{radivilova2018decrypting}),
and quality of service management (insuring resources are 
appropriate for traffic \cite{orsolic2017machine}).
Traffic analysis can be done using:
port structure (services being used) \cite{dainotti2012issues},
deep packet inspection (signatures in payloads) \cite{dainotti2012issues},
data analytics (per-packet or per-flow) \cite{cao2014survey}, or
behavioral classification (communication pattern graphs) \cite{cao2014survey}.

To overcome the weaknesses associated with the port-based and 
payload-based approaches, data analytics has been proposed as a 
solution for network traffic 
classification \cite{alshammari2009machine,erman2006traffic}.
In particular, data-analytic approaches can detect novel traffic
using anomaly detection \cite{zhang2015robust}. 

Draper-Gil et al. \cite{draper2016characterization} generated
a dataset used by several researchers to predict traffic classes.
It contains browsing, email, chat, streaming, file transfer,
VoIP, and P2P traffic; each inside and outside of a VPN tunnel.
However, each window contains only a single kind of traffic.
Draper-Gil used k-Nearest Neighbor and decision tree predictors;
Saber et al. \cite{saber2018encrypted} used under- and over-sampling,
PCA, SVMs;
Lotfollahi et al. \cite{lotfollahi2017deep} used stacked autoencoders
and convolutional neural networks; and
Vu et al. \cite{vu2018time} used LSTMs.
The best models on this dataset achieve
F1 scores up to 0.98 on the single-class prediction problem. 

Zhang et al. \cite{zhang2011inferring}, in the closest work to
ours, classify categories of seven types of activities: 
Browsing, Chat, Online Gaming, Downloading, Uploading, 
Online Video, and BitTorrent, using windows of different sizes.
However, each window contains at most two simultaneous kinds of
traffic.
They achieve around 80\% prediction accuracy for 5-second windows and 
over 90\% accuracy for 1-minute traffic windows for the single class problem. 
Their accuracy decreases for some classes when there are two concurrent activities.

A finer-grained, and harder, version of the problem is to detect not only the
kind of traffic but which application is being used.
Taylor et al. \cite{taylor2016appscanner} create an ``Appscanner'' tool to 
recognize smartphone apps in encrypted traffic. 
Their work was later improved by Taylor et al. \cite{taylor2018robust} using
bursts, packets grouped within a time window. 
They analyze one second bursts to deliver near real-time prediction for
110 out of 200 most popular free Google Play applications. 
Alan and Kaur \cite{alan2016can} reduce computational complexity by 
analyzing only the first 64 packets to identify the application. 
Saltaformaggio et al. \cite{saltaformaggio2016eavesdropping} achieve an average of 
78\% precision and 76\% recall using KMeans and multi-class SVM and properties
that could be captured by eavesdropping. 
In addition, they also paid attention to how easy it was to discover user properties from
the traffic, and revealed some major privacy issues. 
Application identification is also possible in encrypted tunnels. 
Lotfollahi et al. \cite{lotfollahi2017deep} and Yamansavascilar et al. \cite{yamansavascilar2017application} run 
their experiments on the same VPN dataset by Draper-Gil et al. \cite{draper2016characterization} 
to detect application instead of the broad traffic category. 

A related problem is to detect what actions users are doing on their
devices, that is behavioral detection.
Conti et al. \cite{conti2016analyzing} analyze encrypted traffic from 
Android devices to discover user behaviors such as sending a new message 
with Gmail or opening the Dropbox app. 
Their study included seven Android applications: 
Facebook, Gmail, Twitter, Tumblr, Dropbox, Google+ and Evernote
and they simulated user actions to obtain signatures of flows generated
by use of these apps.
Coull and Dyer analyze user behaviors in Apple iMessage \cite{coull2014traffic}
and showed that users' actions, as well as language and length of messages exchanged,
can be predicted.
Park and Kim \cite{park2015encryption} predict KakaoTalk's eleven behaviors 
in encrypted traffic.
Liu et al. \cite{liu2017effective} predict whether a user is using Wechat, 
WhatsApp and Facebook and behaviors such as voice calling, 
video calling or picture sharing.
Dubin et al. \cite{dubin2017know} trained a classifier to
predict the most popular YouTube videos titles, basing on 
encrypted traffic statistics, notably the bits per peak.

\section{Approach}

A real-world system was used to collect RDP traffic data about five
classes of traffic.
Derived data were computed from the base traffic data using
Discrete Cosine Transform, singular value decomposition, and
independent component analysis.
A series of predictive models were then built and their performance
assessed using standard measures as well as a custom measure
designed for the problem.

\subsection{System setup}

The Remote Desktop Protocol is designed to let a user at a
client interact with a host as if sitting at that host.
It provides the full interactivity of using the mouse, viewing the
screen, and using the keyboard.

The client workstation, called Workstation01,
is a 6-core Intel Xeon processor, 36 GB memory, 
2 TB storage and Network Interface Card connected to a
Local Area Network with an Internet gateway, 
running Microsoft Windows 10 Professional.
For data analysis, Java Runtime Environment 1.8 was installed to run 
CIC FlowMeter v4.0 for attribute extraction,
and Wireshark to capture RDP network traffic into .pcap files.
As a Remote Desktop client, Workstation01 is configured to mount local 
drives, allow clipboard and runs in resolution 1366x768 with full user 
desktop experience. 

Hyper-V service is enabled on Workstation01 for 
virtualization and there are two virtual machines running on it:  
Win01 and Centos-Miner.
Win01 VM is a plain installation of a Windows 10 Professional 1809 
virtual machine that acts as a Remote Desktop host. 
The VM is configured with 4 virtual cores and 8 GB of RAM. 
It has a static IP, Network Level Authentication, and a Windows firewall
with an exception allowing incoming RDP traffic.
This setup ensures that Workstation01 can connect freely to Win01 VM over 
both TCP and UDP protocols on port 3389. 

CentOS-Miner is a CentOS 7.5 VM running on top of Workstation-01 Hyper-V. 
This Linux virtual machine has a Bash environment and Tshark installed and 
it is used for data exploration and additional attribute extraction. 

A physically remote VM, CAC01, is hosted at Queen's University's 
Centre for Advanced Computing. 
The configuration is exactly the same as for Win01 VM except
that it runs Windows 10 Education,
and resides behind a firewall with NAT. 
CAC-01 has 4 virtual cores and 4 GB of RAM. 

Traffic to Win01 is on the same IP subnet as the client, and
only encounters the built-in Windows firewall on Win01 itself.
Traffic to CAC01 passes through an Internet connection, the
Queen's University campus network, and a physical firewall that
accepts connections on port 13389 and forwards them to CAC01 on
port 3389.
RDP uses TCP only for traffic outside a subnet, and a mixture of TCP
and UDP for traffic within a subnet, so two different predictive models
had to be developed.

\begin{figure*}
\centering
 \includegraphics[width=0.8\textwidth]{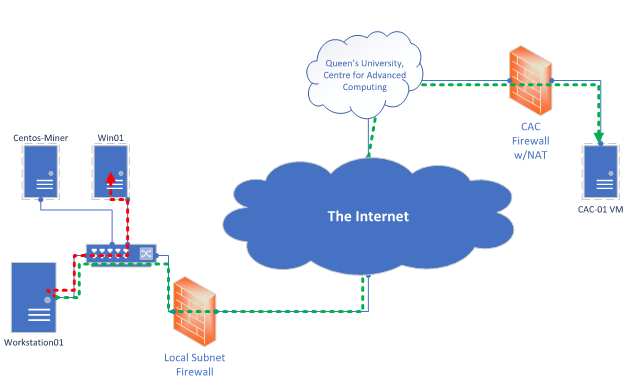}
 \caption{Network architecture diagram}
 \label{fig:31png}
\end{figure*}

Figure \ref{fig:31png} shows the network setup with all the 
Virtual Machines used for data generation. 
The red path shows the traffic path for the local subnet scenario 
and the green path represents the distant scenario, traversing the Internet.
 
\subsection{Data Collection}

The Remote Desktop traffic attributes were captured for 30 second windows
of one or more of the following activity classes:
\begin{itemize}
\item Download -- file download from host to client;
\item Browsing -- using a browser (Firefox and Chrome) on the host, 
driven from the client;
\item Notepad -- editing on the host from the client (typing for more than
80\% of the time window);
\item YouTube -- playing a video on the host (Youtube or mp4s), 
viewed on the client; and
\item Clipboard -- copying content from remote system to local using the
clipboard mechanism.
\end{itemize}

Samples may be pure -- a single activity for 30 seconds -- or mixed, with
up to four activities simultaneously in a 30-second window.

Attribute collection is performed by CIC Flow Meter and Tshark. 
The former is a tool developed by University of New Brunswick's 
Canadian Institute for Cybersecurity \cite{lashkari2017characterization}. 
CIC Flow Meter extracts a predefined number of attributes for each network traffic 
conversation from a .pcap file.
There will be one conversation for TCP only traffic samples, and two conversations 
for TCP and UDP traffic samples. Predictive attributes include properties 
such as packet lengths, inter-arrival times and TCP SYN flag counts.

\begin{figure}
 \centering
 \includegraphics[width=0.75\columnwidth]{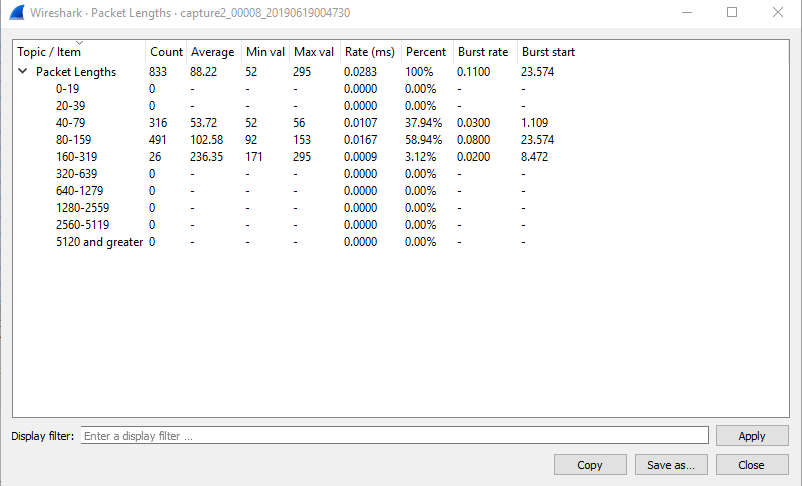}
 \caption{Wireshark -- packet length statistics}
 \label{fig:32png}
\end{figure}

A Bash script extracts additional meaningful predictive attributes
from the packet-level attributes of Tshark.
The additional attributes include Packet Length Statistics 
from Wireshark, as shown in Figure \ref{fig:32png}, 
and include attributes discovered during exploratory data analysis.
All of the data is preprocessed to insure that forward always
refers to traffic from local to remote, and backward the converse,
based on IP addresses.

Several derived attributes of the measured traffic attributes were
also computed.
The Discrete Cosine Transform expresses a sequence as a sum of cosine
functions.
It has often been observed that a single component of the
DCT integrates the variability within a sequence
\cite{makhoul1980fast}. 
We compute this value for each row of the dataset, that is for the
collected traffic attributes of each
traffic flow.

Singular Value Decomposition is an affine transformation of a vector
space into a form where the greatest variability is captured by the
first few axes of the transformed space.
Truncation of the SVD amounts to a projection of a high-dimensional
space into a low-dimensional one in a way that preserves maximal
variation.
We apply SVD to the matrix of windows $\times$ traffic attributes
and keep the first 20 columns of the left singular vector matrix.

Independent Component Analysis (ICA) \cite{hyvarinen1999fast} is 
another matrix transformation that decomposes a matrix into statistically
independent components.
We apply ICA to the windows $\times$ traffic attributes and select
the 20 columns of the left decomposition matrix\footnote{There are
a number of sophisticated ways to select the most important columns,
for example those with maximal kurtosis, but we retain the first 20
and leave it to attribute selection to find the best.}.

\begin{table}[]
\scriptsize
\centering
\caption{Attributes}
\label{tab:Predictive_Attributes}
\begin{tabular}{@{}|l|l|l|l|@{}}
\hline
\multicolumn{4}{|c|}{\textbf{Attributes}}    \\ \hline
\begin{tabular}[c]{@{}l@{}}ACK Flag Cnt\\  Active Max\\  Active Mean\\  Active Min\\  Active Std\\  Bwd Blk Rate Avg\\  Bwd Byts/b Avg\\  Bwd Header Len\\  Bwd IAT Max\\  Bwd IAT Mean\\  Bwd IAT Min\\  Bwd IAT Std\\  Bwd IAT Tot\\  Bwd PSH Flags\\  Bwd Pkt Len Max\\  Bwd Pkt Len Mean\\  Bwd Pkt Len Min\\  Bwd Pkt Len Std\\  Bwd Pkts/b Avg\\  Bwd Pkts/s\\  Bwd Seg Size Avg\\  Bwd URG Flags\\  CWE Flag Count\\  ECE Flag Cnt\\  FIN Flag Cnt\\  Flow Byts/s\\  Flow Duration\\  Flow IAT Max\\  Flow IAT Mean\\  Flow IAT Min\\  Flow IAT Std\\  Flow Pkts/s\end{tabular} & \begin{tabular}[c]{@{}l@{}}Fwd Blk Rate Avg\\  Fwd Byts/b Avg\\  Fwd Header Len\\  Fwd IAT Max\\  Fwd IAT Mean\\  Fwd IAT Min\\  Fwd IAT Std\\  Fwd IAT Tot\\  Fwd PSH Flags\\  Fwd Pkt Len Max\\  Fwd Pkt Len Mean \\ Fwd Pkt Len Min\\  Fwd Pkt Len Std\\  Fwd Pkts/b Avg\\  Fwd Pkts/s\\  Fwd Seg Size Avg\\  Fwd URG Flags\\  Idle Max\\  Idle Mean\\  Idle Min\\  Idle Std\\  Init Bwd Win Byts\\  Init Fwd Win Byts\\  PSH Flag Cnt\\  Pkt Len Max\\  Pkt Len Mean\\  Pkt Len Min\\  Pkt Len Std\\  Pkt Len Var\\  Pkt Size Avg\\  RST Flag Cnt\\  SYN Flag Cnt \end{tabular} & \begin{tabular}[c]{@{}l@{}}Subflow Bwd Byts\\  Subflow Bwd Pkts\\  Subflow Fwd Byts\\  Subflow Fwd Pkts\\  Tot Bwd Pkts\\  Tot Fwd Pkts\\  TotLen Bwd Pkts\\  TotLen Fwd Pkts\\  URG Flag Cnt\\  FwdFrame91-93\\  FwdFrame80-91\\  FwdFrame90-94\\  FwdFrame96-98\\  FwdFrame103-105\\  FwdFrame1280-2559\\  BwdFrame40-79\\  BwdFrame80-159\\  BwdFrame160-319\\  BwdFrame320-639\\  BwdFrame640-1279\\  BwdFrame1280-2559\\  BwdPUSH\\  FwdPUSH\\  dct\_col\\  svd0\\  svd1\\  svd2\\  svd3\\  svd4\\  svd5\\  svd6\\  svd7\end{tabular} & \begin{tabular}[c]{@{}l@{}}svd8\\  svd9\\  svd10\\  svd11\\  svd12\\  svd13\\  svd14\\  svd15\\  svd16\\  svd17\\  svd18\\  svd19\\  ica0\\  ica1\\  ica2\\  ica3\\  ica4\\  ica5\\  ica6\\  ica7\\  ica8\\  ica9\\  ica10\\  ica11\\  ica12\\  ica13\\  ica14\\  ica15\\  ica16\\  ica17\\  ica18\\  ica19\end{tabular}
 \\ \hline
\end{tabular}
\end{table}

The final predictive attribute set is shown in 
Table \ref{tab:Predictive_Attributes}.

Shapley Values calculate the contributions to a final result 
made by individual players in a game theory context \cite{liben2012computing}. 
They have been successfully applied to attribute selection in
predictors because of the development of fast approximation
algorithms that avoid the implicit exponential number of
attribute combinations to be evaluated.
The two most popular Shapley value explainers are 
TreeExplainer \cite{lundberg2019explainable}, for tree-based predictors, 
and DeepExplainer, for neural networks predictors.

We use Shapley values for attribute selection.
The selection process is run independently five times, 
each time for a different class.
Shapley Values were computed with two different classification techniques 
as the backend: Neural Network (Deep Explainer) and 
Extreme Gradient Boosting (XGBoost). 

\subsection{Individual Techniques and the Ensemble Predictor}

For predictors, we use k-Nearest Neighbors, 
Support Vector Machines \cite{berwick2003idiot},
decision trees,
random forests \cite{breiman2001random},
Adaboost \cite{freund1997decision}, 
XGBoost \cite{chen2016xgboost},
and multilayer perceptrons.

Each of the prediction techniques was run with 10-fold cross validation,
with binary class labels (traffic class present or not).
The final ensemble model consists of the top three 
most effective classifiers for each traffic class, 
chosen to maximize the precision, so that the ensemble classifier can 
determine with a high degree of 
certainty whenever a specific kind of traffic is present. 
A specific sample is predicted to contain a specific class of traffic
when at least two of the three classifiers predict that class.

The output of the ensemble classifier is 5 binary values, predicting
the presence of each of the 5 kinds of traffic in the sample.
Three different performance measures are calculated:
\begin{itemize}
\item The precision, recall and F1 score for each class;
\item A confusion matrix for each class;
\item An ensemble score function designed for the problem domain.
\end{itemize}
The ensemble score function assigns points to each record as follows:
\begin{itemize}
\item For a true negative, no points;
\item For a true positive, add one point;
\item For a false positive, subtract two points;
\item For a false negative, no points.
\end{itemize}
For instance, if the model predicts download and clipboard, where the 
traffic actually contained download and browsing, it will assign $+1$ 
point for download, 0 points for browsing as it was a false negative, 
0 points for Notepad and YouTube as they were true negatives, 
and $-2$ for false positive on clipboard. 
The aggregate of scores calculated on all test set rows is divided by 
the sum of all positive labels in the test set and multiplied by 100. 
The scoring function severely penalizes the ensemble whenever
it predicts a traffic class that is not actually present.
This insures that the model is reliable for predicting
behavior.

\subsection{Dataset}

The dataset consists of 2160 30-second data samples. 
The information relating to number of samples 
is summarized in Table \ref{tab:TCPsamples} for the CAC-01 
(Remote VM, TCP transport) and Table \ref{tab:UDPsamples} for Win01 
(Local subnet VM, TCP and UDP transport) traffic respectively. 
For simplicity, we will refer to the first kind of traffic as TCP and 
to the latter as UDP.

\begin{table*}[h]
\footnotesize
\centering
\caption{Total TCP Samples Count}
\label{tab:TCPsamples}
\begin{tabular}{@{}|l|l|l|l|l|l|@{}}
\hline
\textbf{Num samples} & \textbf{Download?} & \textbf{Browsing?} & \textbf{Notepad?} & \textbf{YouTube?} & \textbf{Clipboard?} \\\hline
240            & 1         & 0         & 0         & 0         & 0         \\\hline
240            & 0         & 1         & 0         & 0         & 0         \\\hline
239            & 0         & 0         & 1         & 0         & 0         \\\hline
243            & 0         & 0         & 0         & 1         & 0         \\\hline
120            & 0         & 0         & 0         & 0         & 1         \\\hline
74             & 0         & 1         & 0         & 0         & 1         \\\hline
43             & 1         & 1         & 0         & 0         & 0         \\\hline
25             & 1         & 0         & 1         & 0         & 1         \\\hline
22             & 1         & 0         & 1         & 0         & 0         \\\hline
62             & 0         & 0         & 1         & 1         & 0         \\\hline
27             & 1         & 0         & 0         & 1         & 0         \\\hline
63             & 0         & 1         & 0         & 1         & 0         \\\hline
22             & 0         & 0         & 1         & 0         & 1         \\\hline
15             & 1         & 1         & 0         & 0         & 1         \\\hline
21             & 1         & 0         & 1         & 1         & 1         \\\hline
1456            & \multicolumn{5}{l|}{TOTAL}                                    \\\hline
\end{tabular}
\end{table*}

\begin{table*}[h]
\footnotesize
\centering
\caption{Total UDP Samples Count}
\label{tab:UDPsamples}
\begin{tabular}{@{}|l|l|l|l|l|l|@{}}
\hline
\textbf{Num samples} & \textbf{Download?} & \textbf{Browsing?} & \textbf{Notepad?} & \textbf{YouTube?} & \textbf{Clipboard?} \\\hline
103            & 1         & 0         & 0         & 0         & 0         \\\hline
92             & 0         & 1         & 0         & 0         & 0         \\\hline
100            & 0         & 0         & 1         & 0         & 0         \\\hline
105            & 0         & 0         & 0         & 1         & 0         \\\hline
100            & 0         & 0         & 0         & 0         & 1         \\\hline
42             & 0         & 1         & 0         & 0         & 1         \\\hline
44             & 0         & 1         & 0         & 1         & 0         \\\hline
42             & 1         & 0         & 1         & 0         & 0         \\\hline
37             & 1         & 0         & 1         & 0         & 1         \\\hline
39             & 1         & 0         & 0         & 1         & 0         \\\hline
704            & \multicolumn{5}{l|}{TOTAL}                                    \\\hline
\end{tabular}
\end{table*}

When TCP is the only protocol used, it is easier to make accurate predictions. 
When UDP is added, the problem complexity grows because there
are at least two conversations -- one for the TCP stream and 
one for the UDP stream, in the same Remote Desktop session. 
UDP seems to be used for bulk transfers while TCP is used to send user 
inputs, such as keystrokes and mouse movements, 
and perhaps RDP session management details.

Every keystroke that is sent from local system to the remote system over 
Remote Desktop is carried by two 92-byte TCP frames in the forward direction. 
As a result, the total number of 92-byte frames in a window reveals
how many keystrokes have been sent (except that actions such as pressing and 
holding a Shift key generates 92-byte frames, until the key is released). 

The remote system also responds with packets that are revealing. 92-byte
upward packets produce PSH-flagged packets whose payload size correlates
with the visual change to the screen of the character echo -- the more
pixels changed, the larger the payload\footnote{There are hints that
what is being returned is the delta of the character being displayed.}.
This suggests a potential attack against passwords, since the echoed character
is typically not the character sent but a filler character.
As a result, entering a password may produce an easily
detectable signature.

Mouse movements generate packets of either 97 or 104 bytes, with
97-byte packets associated with mouse clicks.
Observing the number of such packets allows mouse activity to be
estimated; and the mixing with 92-byte packets allows even finer
grained estimates -- for example, the size of form data might be
estimated by observing how many characters are typed in between
mouse clicks and small mouse movements.

\section{Results}

Attribute ranking is computed for ten cases:
5 TCP activities and 5 UDP activities, and for each both the tree-based
and neural-net based Shapley value computation.
This results in twenty ranked lists of attribute significance.
Tree based Shapley values also provide information about whether an
attribute is associated with one class label or the other, and how strongly.
These are shown as red and blue points in plots.

\begin{figure}
 \centering
 \includegraphics[width=0.85\columnwidth]{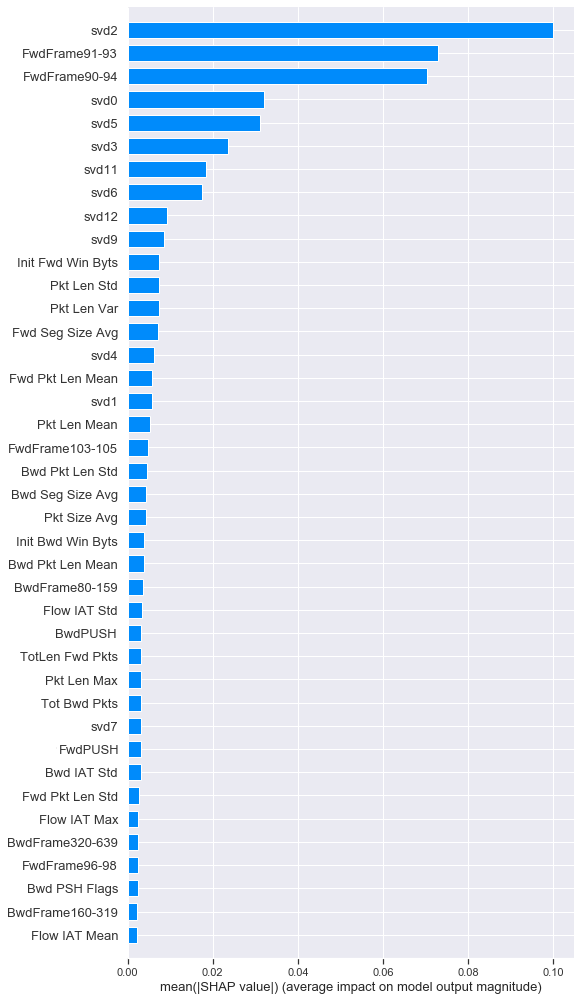}
 \caption{TCP Attribute Ranking -- Notepad, Deep Explainer}
 \label{fig:TCP2DE}
\end{figure}

\begin{figure}
 \centering
 \includegraphics[width=0.85\columnwidth]{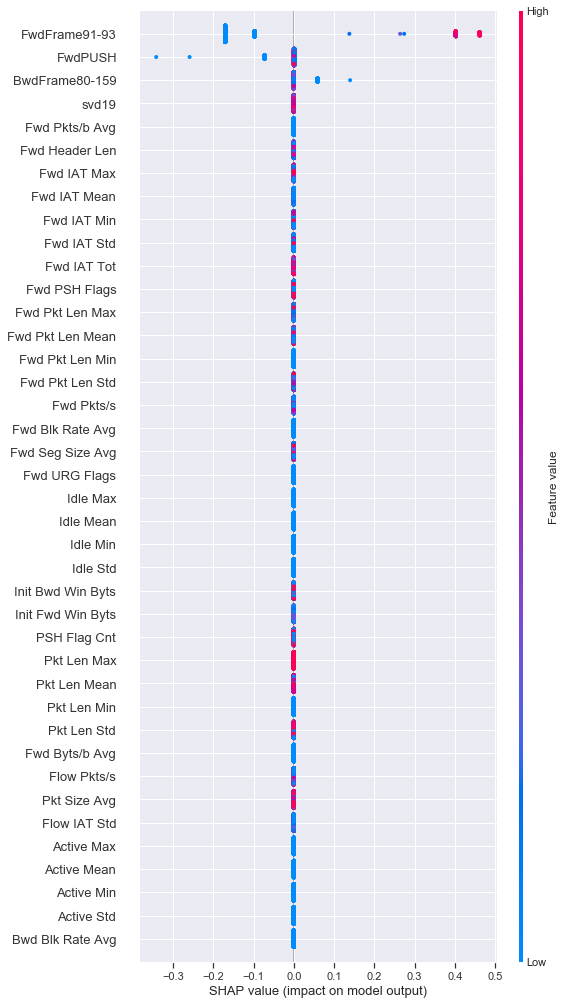}
 \caption{TCP Attribute Ranking -- Notepad, XGBoost Explainer}
 \label{fig:TCP2XGB}
\end{figure}

\begin{figure}
 \centering
 \includegraphics[width=0.85\columnwidth]{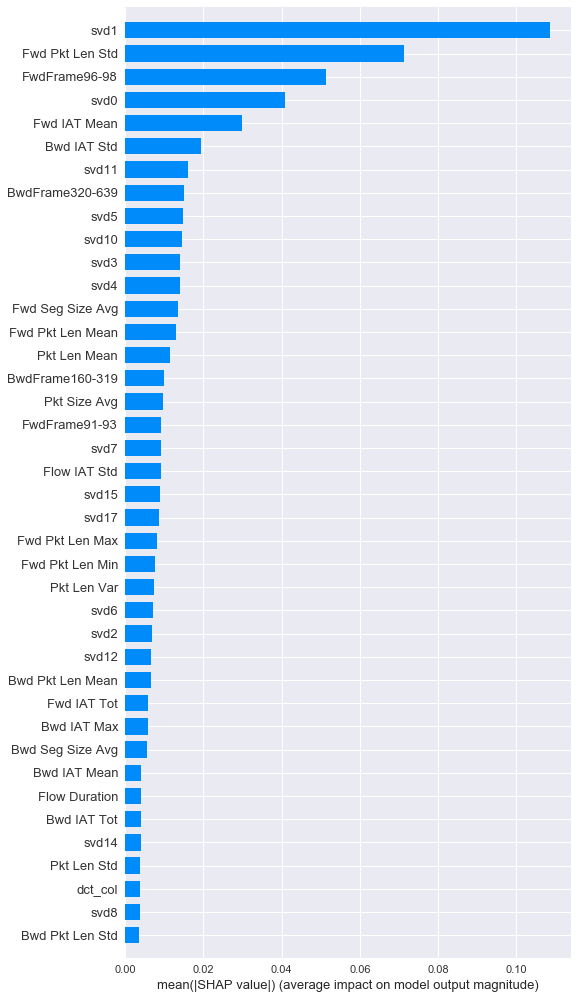}
 \caption{UDP Attribute Ranking -- Browsing, Deep Explainer}
 \label{fig:UDP1DE}
\end{figure}

\begin{figure}
 \centering
 \includegraphics[width=0.85\columnwidth]{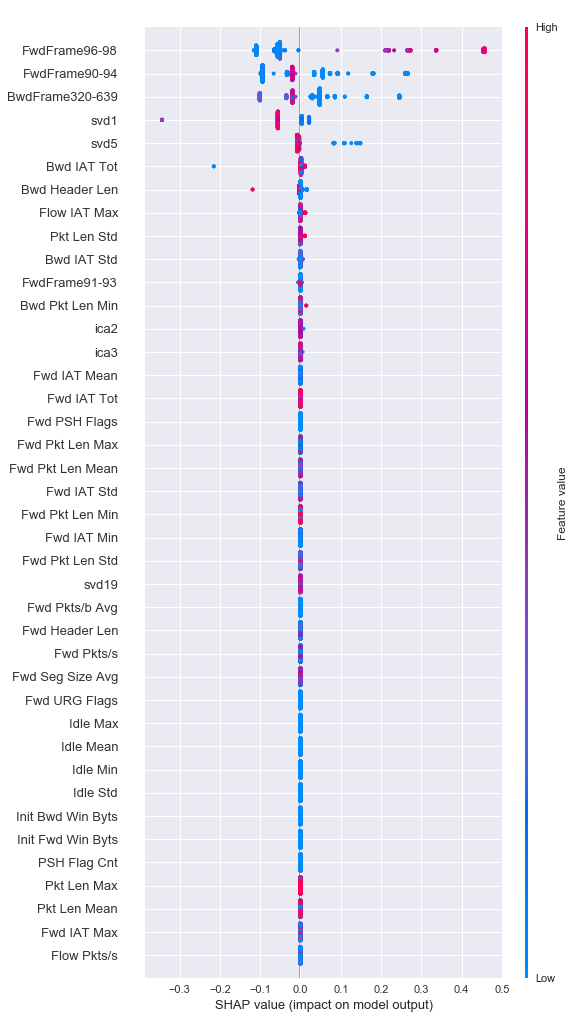}
 \caption{UDP Attribute Ranking -- Browsing, XGBoost Explainer}
 \label{fig:UDP1XGB}
\end{figure}

The most predictive attributes are fairly consistent using both the tree-based
and network-based Shapley values. 
The 92- and 104-byte attributes rank highly, as do many of the SVD
components, suggesting that the original attributes tend to capture
traffic properties that are strongly correlated.

\begin{table*}[h]
\footnotesize
\centering
\caption{Selected Attribute Sets for TCP Classes}
\label{tab:tcp-feats}
\begin{tabular}{|c|c|c|c|c|}
\hline
\textbf{TCP Download}

& \textbf{TCP Browsing}                                                                                                                                    
& \textbf{TCP Notepad}                                                                   
& \textbf{TCP YouTube}                                                                                                 
& \textbf{TCP Clipboard}                                                                                                                           \\ \hline
\begin{tabular}[t]{@{}c@{}}svd0 \\ BwdFrame80-159 \\ BwdFrame320-639 \\ Init Fwd Win Byts \\ Tot Fwd Pkts \\ BwdFrame640-1279 \\ Fwd IAT Mean \\ Pkt Len Std \\ Bwd IAT Max \\ ica19 \\ Bwd Pkt Len Mean \\ Flow Byts/s\end{tabular} & 
\begin{tabular}[t]{@{}c@{}}FwdFrame103-105 \\ BwdFrame320-639 \\ Fwd Pkt Len Mean \\ Fwd Seg Size Avg \\ svd8 \\ BwdPUSH \\ FwdFrame96-98 \\ FwdPUSH \\ BwdFrame1280-2559 \\ ica7 \\ FwdFrame91-93 \\ Init Fwd Win Byts \\ Bwd Pkt Len Mean\end{tabular} & 
\begin{tabular}[t]{@{}c@{}}svd2 \\ FwdFrame91-93 \\ svd0 \\ FwdPUSH \\ BwdFrame80-159\end{tabular} & 
\begin{tabular}[t]{@{}c@{}}BwdFrame320-639 \\ svd1 \\ svd6 \\ svd0 \\ svd9 \\ FwdPUSH \\ TotLen Fwd Pkts \\ Flow IAT Std \\ BwdFrame640-1279\end{tabular} & 
\begin{tabular}[t]{@{}c@{}}FwdPUSH \\ svd1 \\ svd0 \\ Subflow Bwd Byts \\ ica7 \\ FwdFrame91-93 \\ FwdFrame103-105 \\ svd6 \\ ica11 \\ svd11 \\ ica5 \\ Fwd Pkt Len Mean \\ BwdFrame320-639 \\ svd8 \\ Pkt Len Std\end{tabular} \\ \hline
\end{tabular}
\end{table*}

\begin{table*}[h]
\footnotesize
\centering
\caption{Selected Attribute Sets for UDP Classes}
\label{tab:udp-feats}
\begin{tabular}{|c|c|c|c|c|}
\hline
\textbf{UDP Download}     & \textbf{UDP Browsing}                                                   & \textbf{UDP Notepad}                                                     & \textbf{UDP YouTube}                                                                                                                                                               & \textbf{UDP Clipboard}                                                                                                                       \\ \hline
\begin{tabular}[t]{@{}c@{}}svd0 \\  svd1\\  svd2\\  Fwd Pkt Len Std\\  Bwd Header Len\end{tabular} & 
\begin{tabular}[t]{@{}c@{}}svd1\\  Fwd Pkt Len Std\\  FwdFrame96-98\\  svd0\\  svd5\\  FwdFrame90-94\\  BwdFrame320-639\end{tabular} & 
\begin{tabular}[t]{@{}c@{}}FwdFrame91-93 \\  BwdFrame40-79 \\  svd2\\  Fwd IAT Min\\  svd1\\  Bwd IAT Mean\end{tabular} & 
\begin{tabular}[t]{@{}c@{}}svd1 \\  svd0 \\  svd3\\  svd5 \\  BwdFrame80-159\\  FwdFrame91-93\\  Flow IAT Max\\  Bwd IAT Max\\  Fwd Pkt Len Mean\\  Fwd Seg Size Avg\\  BwdFrame320-639\\  Bwd IAT Std\\  Fwd IAT Max\\  svd4\\  Bwd Header Len\\  ica8\\  Bwd Pkt Len Max\\  Fwd IAT Tot\\  Bwd IAT Tot\end{tabular} & 
\begin{tabular}[t]{@{}c@{}}svd2\\  Bwd IAT Std\\  Fwd IAT Mean\\  FwdFrame103-105\\  Flow IAT Mean\\  Flow IAT Std\\  BwdFrame40-79\\  BwdFrame320-639\\  Bwd IAT Mean\\  FwdFrame91-93\\  Pkt Len Mean\\  FwdFrame90-94\\  Fwd IAT Min\end{tabular} \\ \hline
\end{tabular}
\end{table*}

Table~\ref{tab:tcp-feats} shows the best attributes for the
TCP traffic classes and Table~\ref{tab:udp-feats} shows the best attributes 
for the UDP traffic classes.

\begin{table*}[h]
\centering
\small
\caption{TCP Traffic Individual Techniques -- Accuracy and Standard Deviation}
\label{tab:TCP_IndividualTechniques}
\begin{tabular}{|c|c|c|c|c|c|}
\hline
\textbf{Technique} & \textbf{Download} & \textbf{Browsing} & \textbf{Notepad} & \textbf{YouTube} & \textbf{Clipboard} \\ \hline
\textbf{NN} & \begin{tabular}[c]{@{}c@{}}99.66\% \\ ($\pm$ 0.46\%)\end{tabular} & \begin{tabular}[c]{@{}c@{}}98.62\% \\ ($\pm$ 1.27\%)\end{tabular} & \begin{tabular}[c]{@{}c@{}}99.66\% \\ ($\pm$ 0.46\%)\end{tabular} & \begin{tabular}[c]{@{}c@{}}99.59\%\\ ($\pm$ 0.46\%)\end{tabular} & \begin{tabular}[c]{@{}c@{}}95.88\%\\ ($\pm$ 1.23\%)\end{tabular} \\ \hline
\textbf{RF} & \begin{tabular}[c]{@{}c@{}}99.18\%\\ ($\pm$ 0.79\%)\end{tabular} & \begin{tabular}[c]{@{}c@{}}99.38\% \\ ($\pm$ 0.57\%)\end{tabular} & \begin{tabular}[c]{@{}c@{}}99.86\%\\ ($\pm$ 0.27\%)\end{tabular} & \begin{tabular}[c]{@{}c@{}}99.59\% \\ ($\pm$ 0.46\%)\end{tabular} & \begin{tabular}[c]{@{}c@{}}98.69\%\\ ($\pm$ 0.84\%)\end{tabular} \\ \hline
\textbf{XGB} & \begin{tabular}[c]{@{}c@{}}99.52\%\\ ($\pm$ 0.69\%)\end{tabular} & \begin{tabular}[c]{@{}c@{}}99.59\% \\ ($\pm$ 0.62\%)\end{tabular} & \begin{tabular}[c]{@{}c@{}}99.72\% \\ ($\pm$ 0.46\%)\end{tabular} & \begin{tabular}[c]{@{}c@{}}99.59\% \\ ($\pm$ 0.45\%)\end{tabular} & \begin{tabular}[c]{@{}c@{}}98.56\%\\ ($\pm$ 1.12\%)\end{tabular} \\ \hline
\textbf{SVM} & \begin{tabular}[c]{@{}c@{}}99.73\%\\ ($\pm$ 0.34\%)\end{tabular} & \begin{tabular}[c]{@{}c@{}}99.24\%\\ ($\pm$ 0.72\%)\end{tabular} & \begin{tabular}[c]{@{}c@{}}99.86\%\\ ($\pm$ 0.28\%)\end{tabular} & \begin{tabular}[c]{@{}c@{}}99.59\% \\ ($\pm$ 0.45\%)\end{tabular} & \begin{tabular}[c]{@{}c@{}}95.47\% \\ ($\pm$ 2.60\%)\end{tabular} \\ \hline
\textbf{Ada} & \begin{tabular}[c]{@{}c@{}}99.66\%\\ ($\pm$ 0.46\%)\end{tabular} & \begin{tabular}[c]{@{}c@{}}99.25\% \\ ($\pm$ 0.57\%)\end{tabular} & \begin{tabular}[c]{@{}c@{}}99.86\%\\ ($\pm$ 0.28\%)\end{tabular} & \begin{tabular}[c]{@{}c@{}}99.52\%\\ ($\pm$ 0.81\%)\end{tabular} & \begin{tabular}[c]{@{}c@{}}98.49\%\\ ($\pm$ 0.67\%)\end{tabular} \\ \hline
\textbf{KNN} & \begin{tabular}[c]{@{}c@{}}99.45\%\\ ($\pm$ 0.51\%)\end{tabular} & \begin{tabular}[c]{@{}c@{}}99.04\%\\ ($\pm$ 0.70\%)\end{tabular} & \begin{tabular}[c]{@{}c@{}}99.79\%\\ ($\pm$ 0.31\%)\end{tabular} & \begin{tabular}[c]{@{}c@{}}99.31\%\\ ($\pm$ 0.68\%)\end{tabular} & \begin{tabular}[c]{@{}c@{}}94.71\%\\ ($\pm$ 1.81\%)\end{tabular} \\ \hline
\textbf{DTC} & \begin{tabular}[c]{@{}c@{}}98.63\%\\ ($\pm$ 0.86\%)\end{tabular} & \begin{tabular}[c]{@{}c@{}}98.83\%\\ ($\pm$ 0.75\%)\end{tabular} & \begin{tabular}[c]{@{}c@{}}99.93\% \\ ($\pm$ 0.21\%)\end{tabular} & \begin{tabular}[c]{@{}c@{}}99.52\%\\ ($\pm$ 0.69\%)\end{tabular} & \begin{tabular}[c]{@{}c@{}}96.91\%\\ ($\pm$ 1.20\%)\end{tabular} \\ \hline
\end{tabular}
\end{table*}

\begin{table*}[h]
\centering
\small
\caption{UDP Traffic Individual Techniques -- Accuracy and Standard Deviation}
\label{tab:UDP_IndividualTechniques}
\begin{tabular}{|c|c|c|c|c|c|}
\hline
\textbf{Technique} & \textbf{Download} & \textbf{Browsing} & \textbf{Notepad} & \textbf{YouTube} & \textbf{Clipboard} \\ \hline
\textbf{NN} & \begin{tabular}[c]{@{}c@{}}99.86\%\\ ($\pm$ 0.42\%)\end{tabular} & \begin{tabular}[c]{@{}c@{}}97.72\%\\ ($\pm$ 1.93\%)\end{tabular} & \begin{tabular}[c]{@{}c@{}}99.71\% \\ ($\pm$ 0.57\%)\end{tabular} & \begin{tabular}[c]{@{}c@{}}99.86\%\\ ($\pm$ 0.42\%)\end{tabular} & \begin{tabular}[c]{@{}c@{}}96.02\% \\ ($\pm$ 2.09\%)\end{tabular} \\ \hline
\textbf{RF} & \begin{tabular}[c]{@{}c@{}}99.86\%\\ ($\pm$ 0.42\%)\end{tabular} & \begin{tabular}[c]{@{}c@{}}99.29\% \\ ($\pm$ 0.71\%)\end{tabular} & \begin{tabular}[c]{@{}c@{}}99.43\% \\ ($\pm$ 0.69\%)\end{tabular} & \begin{tabular}[c]{@{}c@{}}100.00\% \\ ($\pm$ 0.00\%)\end{tabular} & \begin{tabular}[c]{@{}c@{}}98.15\%\\ ($\pm$ 1.28\%)\end{tabular} \\ \hline
\textbf{XGB} & \begin{tabular}[c]{@{}c@{}}99.57\%\\ ($\pm$ 0.91\%)\end{tabular} & \begin{tabular}[c]{@{}c@{}}99.01\%\\ ($\pm$ 0.90\%)\end{tabular} & \begin{tabular}[c]{@{}c@{}}98.72\%\\ ($\pm$ 1.34\%)\end{tabular} & \begin{tabular}[c]{@{}c@{}}99.15\% \\ ($\pm$ 0.95\%)\end{tabular} & \begin{tabular}[c]{@{}c@{}}97.87\% \\ ($\pm$ 1.30\%)\end{tabular} \\ \hline
\textbf{SVM} & \begin{tabular}[c]{@{}c@{}}99.71\% \\ ($\pm$ 0.57\%)\end{tabular} & \begin{tabular}[c]{@{}c@{}}98.29\%\\ ($\pm$ 1.39\%)\end{tabular} & \begin{tabular}[c]{@{}c@{}}99.72\%\\ ($\pm$ 0.57\%)\end{tabular} & \begin{tabular}[c]{@{}c@{}}99.43\%\\ ($\pm$ 0.69\%)\end{tabular} & \begin{tabular}[c]{@{}c@{}}95.03\%\\ ($\pm$ 2.73\%)\end{tabular} \\ \hline
\textbf{Ada} & \begin{tabular}[c]{@{}c@{}}99.57\%\\ ($\pm$ 0.65\%)\end{tabular} & \begin{tabular}[c]{@{}c@{}}99.01\%\\ ($\pm$ 0.91\%)\end{tabular} & \begin{tabular}[c]{@{}c@{}}99.57\%\\ ($\pm$ 0.65\%)\end{tabular} & \begin{tabular}[c]{@{}c@{}}99.71\% \\ ($\pm$ 0.86\%)\end{tabular} & \begin{tabular}[c]{@{}c@{}}98.44\% \\ ($\pm$ 1.00\%)\end{tabular} \\ \hline
\textbf{KNN} & \begin{tabular}[c]{@{}c@{}}99.86\% \\ ($\pm$ 0.42\%)\end{tabular} & \begin{tabular}[c]{@{}c@{}}98.01\% \\ ($\pm$ 1.45\%)\end{tabular} & \begin{tabular}[c]{@{}c@{}}99.72\%\\ ($\pm$ 0.56\%)\end{tabular} & \begin{tabular}[c]{@{}c@{}}99.01\%\\ ($\pm$ 1.11\%)\end{tabular} & \begin{tabular}[c]{@{}c@{}}94.45\% \\ ($\pm$ 2.17\%)\end{tabular} \\ \hline
\textbf{DTC} & \begin{tabular}[c]{@{}c@{}}99.86\% \\ ($\pm$ 0.42\%)\end{tabular} & \begin{tabular}[c]{@{}c@{}}98.87\% \\ ($\pm$ 1.76\%)\end{tabular} & \begin{tabular}[c]{@{}c@{}}99.58\%\\ ($\pm$ 0.65\%)\end{tabular} & \begin{tabular}[c]{@{}c@{}}98.86\%\\ ($\pm$ 1.06\%)\end{tabular} & \begin{tabular}[c]{@{}c@{}}96.60\% \\ ($\pm$ 1.69\%)\end{tabular} \\ \hline
\end{tabular}
\end{table*}

Table~\ref{tab:TCP_IndividualTechniques} shows the performance summary 
for the predictors run on the TCP traffic and 
Table~\ref{tab:UDP_IndividualTechniques} summarizes performance for the UDP traffic. 
Accuracies are over 10-fold cross validation.

\begin{figure}[h]
 \centering
 \includegraphics[width=0.85\columnwidth]{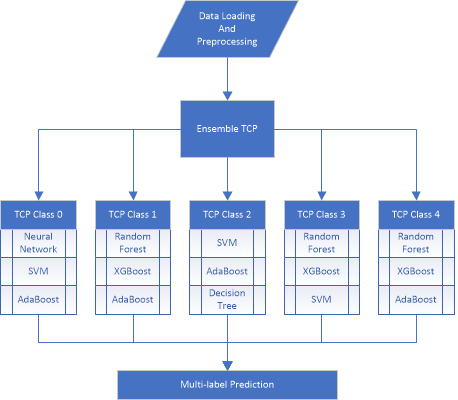}
 \caption{TCP Ensemble Model}
 \label{fig:TCPEnsembleModel}
\end{figure}

\begin{figure}[h]
 \centering
 \includegraphics[width=0.85\columnwidth]{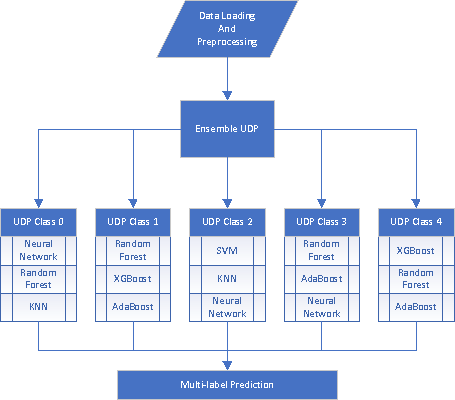}
 \caption{UDP Ensemble Model}
 \label{fig:UDPEnsembleModel}
\end{figure}

For each TCP and UDP traffic class we select the top three best performing 
cross-validated predictors to build two ensemble predictors 
-- the TCP Ensemble and the UDP Ensemble,
shown in Figures~\ref{fig:TCPEnsembleModel} and \ref{fig:UDPEnsembleModel}.

Table \ref{tab:summaryTCPUDP_5fold} provides the
True Positives, False Positives, True Negatives, and False Negatives 
with five-fold cross-validation for the ensemble predictors.
We then calculate Accuracy, Precision, Recall and F1 score.

\begin{table*}[h]
\centering
\caption{Summary of Each TCP and UDP Traffic Class Results over 5-fold cross-validation}
\label{tab:summaryTCPUDP_5fold}
\begin{tabular}{|r|c|c|c|c|c|c|c|c|}
\hline
\multicolumn{1}{|c|}{\textbf{Class}} & \textbf{FP} & \textbf{TP} & \textbf{FN} & \textbf{TN} & \textbf{Accu.} & \textbf{Prec.} & \textbf{Rec.} & \textbf{F1} \\ \hline
TCP Download & 1 & 391 & 2 & 1062 & 99.79 & 99.74 & 99.49 & 99.62 \\ \hline
TCP Browsing & 3 & 434 & 1 & 1018 & 99.73 & 99.31 & 99.77 & 99.54 \\ \hline
TCP Notepad & 0 & 390 & 1 & 1065 & 99.93 & 100.00 & 99.74 & 99.87 \\ \hline
TCP YouTube & 2 & 413 & 3 & 1038 & 99.66 & 99.52 & 99.28 & 99.40 \\ \hline
TCP Clipboard & 5 & 265 & 12 & 1174 & 98.83 & 98.15 & 95.67 & 96.89 \\ \hline
UDP Download & 0 & 221 & 0 & 483 & 100.00 & 100.00 & 100.00 & 100.00 \\ \hline
UDP Browsing & 2 & 175 & 3 & 524 & 99.29 & 98.87 & 98.31 & 98.59 \\ \hline
UDP Notepad & 1 & 178 & 1 & 524 & 99.72 & 99.44 & 99.44 & 99.44 \\ \hline
UDP YouTube & 0 & 188 & 0 & 516 & 100.00 & 100.00 & 100.00 & 100.00 \\ \hline
UDP Clipboard & 4 & 170 & 9 & 521 & 98.15 & 97.70 & 94.97 & 96.32 \\ \hline
\end{tabular}
\end{table*}

Table \ref{tab:ensemble_scores5}, summarizes the single ensemble scores 
obtained for each fold of the TCP and UDP models, along with the average. 
Voting makes a significant contribution towards reducing false positives.

\begin{table}[]
\centering
\caption{Single Ensemble Score Results}
\label{tab:ensemble_scores5}
\begin{tabular}{|c|c|c|}
\hline
\textbf{Fold}                               & \textbf{TCP Ensemble Score}              & \textbf{UDP Ensemble Score}              \\ \hline
1                                     & 98.18                         & 95.36                         \\ \hline
2                                     & 97.86                         & 97.47                         \\ \hline
3                                     & 97.19                          & 97.85                         \\ \hline
4                                     & 97.95                         & 96.22                        \\ \hline
5                                     & 98.13                         & 98.90                         \\ \hline
\begin{tabular}[c]{@{}c@{}}Average \end{tabular} & \begin{tabular}[c]{@{}c@{}}97.86\end{tabular} & \begin{tabular}[c]{@{}c@{}}97.16\end{tabular} \\ \hline
\end{tabular}
\end{table}

For both UDP and TCP traffic, there are similarities when it comes to 
the most commonly misclassified traffic classes.
For both TCP and UDP traffic, the most common misclassification misses
the Clipboard class. This is reflected in the recall scores for the 
UDP Clipboard at only 94.97\% and the TCP Clipboard at 95.67\%, 
while other classes all have recall higher than 98\%.

The most common error in TCP traffic is a mixture of Browsing and 
Clipboard being classified as Browsing only. 
Both Browsing and Clipboard involve mouse movements and mouse clicks, 
both right- and left-clicks, but it seems slightly surprising that these
are difficult to distinguish.

The second most common error occurs when Browsing traffic is classified 
as Browsing and Clipboard, generating a false positive prediction for 
Clipboard. 
The third most common is for the combination of Download, Browsing and Clipboard 
where the Clipboard traffic is missed. 
The fourth most common is when a Download and YouTube mixture is classified 
as Download only. 
Misclassifications are rare the other way -- mixtures of the two rarely 
get classified as YouTube only. Download has a potential to dominate YouTube, 
apparently because it generates significantly more and faster traffic;
and it is possible that backward (remote to local) packet lengths could vary, 
especially when traversing the Internet. 
This could be caused by TCP adjusting the window size for transfers. 
Interestingly, there were no misclassifications relating to the 4-class TCP traffic. 

For UDP traffic, the most common misclassification is again a False 
negative for the Clipboard class. 
Most often, this happens in the mixture of Download, Notepad and Clipboard, 
where Clipboard is missed and traffic is classified as Download and Notepad. 
Both Download and Notepad have many more visible traffic attributes 
than Clipboard does -- the total number of backward bytes 
for Download and 92-byte frames for Notepad. 
There were some other misclassifications such as Browsing and Clipboard 
being identified as Browsing alone and Clipboard or Browsing being 
classified as none of the five classes.

We wondered if Browsing samples from web sites with lots of video
and dynamic content would tend to get misclassified as the YouTube class. 
Such misclassifications are rare and the predictive model is apparently 
resilient to audio and video contents embedded in websites.

\section{Conclusions}

We have shown that, for an encrypted protocol such as RDP,
it is still possible to infer five common categories
of activities with high reliability
from traffic properties that cannot be concealed by encryption.
It is conceivable that some of these predictions could be defeated
by obfuscation in the protocol but protocol designers are
caught between the need to conceal activity and the need to provide
responsiveness.
As we have shown, this has led to a design in which keystrokes,
mouse activity, and visual rendering all leave traces in the
encrypted traffic which could potentially be used for attacks.

A limitation of this model is that it only used traffic between Windows 10 
systems. Different systems, and RDP updates, could conceivably change
traffic structure in a macroscopic way, although the proof of concept
here suggests that only model retraining would be needed.
Other kinds of traffic were not examined: for example, two-way video
such as Zoom or Skype might induce quite different traffic structure.


\end{document}